\documentclass{jpsj2}
\usepackage{epsf}
\usepackage{graphicx}
\usepackage{amsfonts}
\usepackage{dcolumn}%
\usepackage{color,ulem}
\usepackage{bm}

\begin{document}
  
\title{0-$\pi$ transition driven by magnetic proximity effect in a Josephson junction} 

\author{Shin-ichi Hikino$^{1}$ and Seiji Yunoki$^{1,2,3}$}%
\inst{%
$^{1}$Computational Condensed Matter Physics Laboratory, RIKEN ASI, Wako, Saitama 351-0198, Japan \\
$^{2}$Computational Materials Science Research Team, RIKEN Advanced Institute for Computational Science (AICS), 
Kobe, Hyogo 650-0047, Japan\\
$^{3}$Computational Quantum Matter Research Team, RIKEN Center for Emergent Matter Science (CEMS), Wako, Saitama 351-0198, Japan
} 

\date{\today}

\abst{
We theoretically study the Josephson effect in a superconductor/normal metal/superconductor ({\it S}/{\it N}/{\it S}) Josephson junction 
composed of $s$-wave {\it S}s with {\it N} which is sandwiched by two ferromagnetic insulators ({\it F}s), forming a spin valve, 
in the vertical direction of the junction. 
We show that the 0-$\pi$ transition of the Josephson critical current occurs with increasing the thickness of {\it N} along the junction. 
This transition is due to the magnetic proximity effect (MPE) which induces ferromagnetic magnetization in the {\it N}. 
Moreover, we find that, even for fixed thickness of {\it N}, the proposed Josephson junction with the spin valve can be 
switched from $\pi$ to 0 states and vice versa by varying the magnetization configuration (parallel or antiparallel) of two {\it F}s. 
We also examine the effect of spin-orbit scattering on the Josephson critical current and argue that the 0-$\pi$ transition 
found here can be experimentally observed within the current nanofabrication techniques, thus 
indicating a promising potential of this junction as a 0-$\pi$ switching device operated reversibly with varying the magnetic configuration 
in the spin valve by, e.g., applying an external magnetic field. 
Our results not only provide possible applications in superconducting electronics but also suggest the importance of 
a fundamental concept of MPE in nanostructures of multilayer {\it N}/{\it F} systems. 
}


\maketitle 


\section{Introduction}\label{sec:introduction}

The DC Josephson effect has been known for more than 50 years as one of the most fundamental phenomena, 
characterized by DC current flowing without a voltage-drop between two superconductors ({\it S}s) separated by a thin 
insulator ({\it I})~\cite{josephson}.  
It is a macroscopic quantum-mechanical phenomenon in which the superconducting phase coherence is preserved between these two {\it S}s. 
A similar phenomenon occurs in a junction of weakly linked {\it S}s through a normal metal ({\it N}) due to the proximity 
effect between {\it S} and {\it N}~\cite{degennes}.  As in an {\it S}/{\it I}/{\it S} junction, the Josephson critical current in an {\it S}/{\it N}/{\it S} junction monotonically 
decreases with the thickness of {\it N}~\cite{likhalev}.

Because of its fascinating phenomena and potential applications to spintronics, $s$-wave superconductor/ferromagnetic metal ({\it S}/{\it FM}) 
hybrid junctions have been rapidly developed in the last decade~\cite{buzdin-jetp, ryazanov-prl, kontos-prl, robinson, golubov-rmp, buzdin-rmp, bergeret-rmp}. 
Due to the proximity effect between {\it S} and {\it FM}, spin-singlet Cooper pairs (SSCs) penetrate into the {\it FM} in an {\it S}/{\it FM} junction. 
Because of the exchange splitting of the electronic density of states for up- and down-spin electrons, 
the SSC has a finite center of mass momentum and thus the pair amplitude of SSC shows damped oscillatory behavior 
with the thickness of {\it FM}. 
One interesting phenomena induced by the damped oscillatory behavior of the pair amplitude is a $\pi$-state in an {\it S}/{\it FM}/{\it S} 
junction, ferromagnetic Josephson junction (FJJ), where 
the current-phase relation in the Josephson current  is shifted by $\pi$ from that of the ordinary {\it S}/{\it I}/{\it S} or {\it S}/{\it N}/{\it S} 
junctions (called 0-state)~\cite{buzdin-jetp, ryazanov-prl, kontos-prl, robinson, golubov-rmp, buzdin-rmp, bergeret-rmp}. 
It is expected that the $\pi$-state can be used for an element of quantum computing and 
circuit~\cite{yamashita-qbit, bell-apl84, khabipov}. 
However, even if such FJJs are fabricated, it is still difficult to freely control the transition from 0-state to 
$\pi$-state or vice versa (0-$\pi$ transition) in FJJs~\cite{mori-jpsj76}. 

A way to regulate the 0-$\pi$ transition in an {\it S}/{\it N}/{\it S} junction has been theoretically 
proposed~\cite{volkov-prl74, wilhelm-prl81, yip-prb58} and experimentally attained~\cite{baselmans-nature, baselmans-aps}. 
Here an {\it S}/{\it N}/{\it S} junction is constructed in a cross geometry with two additional electrodes attached to 
the {\it N}~\cite{volkov-prl74, wilhelm-prl81, yip-prb58, baselmans-nature, baselmans-aps}. 
Depending on the DC voltage ($V_{\rm{DC}}$) applied to these electrodes, the {\it S}/{\it N}/{\it S} junction becomes 
either 0-state or $\pi$-state. 
In this system, the non-equilibrium Fermi distribution function in the {\it N} due to non-zero $V_{\rm {DC}}$ plays 
the essential role to induce the 0-$\pi$ transition~\cite{volkov-prl74, wilhelm-prl81, yip-prb58, baselmans-nature, baselmans-aps, pothier}. 

The magnetic proximity effect (MPE) is a phenomenon in which finite magnetization is induced in the {\it N} or {\it S} when 
the {\it N} or {\it S} is attached to the {\it FM} or ferromagnetic insulator ({\it F})~\cite{hauser, moodera-prb29, antel-prb60, wilhelm-prl85,hernando-prl88, bergeret-prb69,Lofwander-prl95,hernado-epjb44,cottet-prl107,an-ass257,golikova-prb86}. 
Recently, the MPE has attracted much attention in the research field of spintronics because spintronics devices usually
involve {\it N}/ferromagnet junctions and the magnetic transport is often influenced by the 
MPE~\cite{huang-prl109, pugach-apl101, lu-prb87, lu-prl110, qu-prl110, yang-jap115}. 
For instance, the {\it N} film on ferromagnetic substrates exhibits anisotropic magnetoresistance, which is not observed in the {\it N} 
on nonmagnetic substrates~\cite{huang-prl109}. 
Although there are reports suggesting that the MPE interferes with suitable magnetic transport in spintronics 
devices~\cite{huang-prl109,qu-prl110}, 
intriguing phenomena are also expected when the MPE is actively utilized. 
Here, we propose a simple {\it S}/{\it N}/{\it S} junction where the 0-$\pi$ transition is driven and controlled by the MPE\cite{alidoust-prb89}. 

In this paper, we propose a simple {\it S}/{\it N}/{\it S} Josephson junction composed 
of $s$-wave {\it S}s with {\it N} sandwiched by two {\it F}s, forming a spin valve, in the vertical direction of the junction, and 
study theoretically the Josephson effect. 
We show that the 0-$\pi$ transition occurs with increasing the thickness of {\it N} along the junction, which is due to the MPE inducing 
ferromagnetic magnetization in the {\it N}. 
Moreover, we find that, even for fixed thickness of {\it N}, this {\it S}/{\it N}/{\it S} junction can be switched reversibly from $\pi$-state to 
0-state and vice versa by changing the magnetization configuration of two {\it F}s from parallel to antiparallel configuration. 
The effect of spin-orbit scattering on the Josephson critical current is also examined to argue that the 0-$\pi$ transition found here 
can be experimentally observed. 

The rest of this paper is organized as follows. 
In Sec.~\ref{sec:formulation}, we introduce a simple {\it S}/{\it N}/{\it S} junction with the spin valve structure formed by two {\it F}s sandwiching the {\it N}, and 
formulate the Josephson critical current for this junction by solving Usadel equation in the diffusive transport limit. 
In Sec.~\ref{sec:result}, the results of the Josephson critical current are shown as a function of the thickness of {\it N} for different 
magnetization configurations of the two {\it F}s in the spin valve. 
The influence of spin-orbit scattering on the Josephson critical current is also examined. 
Finally, the Josephson critical current is estimated for realistic systems 
in Sec.~\ref{sec:discussion}, 
before summarizing the paper in Sec.~\ref{sec:summary}.

\begin{figure}[t!]
\begin{center}
\includegraphics[width=6cm]{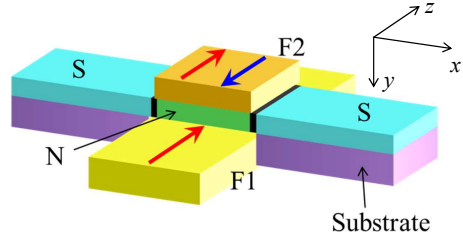}
\caption{ (color online) 
Schematic illustration of the {\it S}/{\it N}/{\it S} junction studied, where the normal metal ({\it N}) is sandwiched by two 
ferromagnetic insulators ({\it F}1 and {\it F}2), forming a spin valve. Arrows in {\it F}1 and {\it F}2 indicate the direction of in-plane ferromagnetic magnetizations. 
While the magnetization in {\it F}1 is fixed, e.g., by exchange bias, the {\it F}2 is assumed to be a free layer in which the magnetization 
direction is controlled by an external magnetic field~\cite{handbook,note1}. 
The thickness of {\it N} in the $x$ direction is $2d_{\rm N }$ and the interfaces between the superconductor ({\it S}) and the {\it N} 
are located at $x=\pm d_{\rm N}$. The uniform magnetization is assumed in 
each {\it F}. 
}
\label{sns-sv}
\end{center}
\end{figure}

\section{Josephson critical current in a Josephson junction with spin valve}\label{sec:formulation}

As depicted in Fig.~\ref{sns-sv}, we consider the Josephson junction composed of two $s$-wave {\it S}s and {\it N}, where the 
{\it N} is $2d_{\rm N}$ thick in the $x$ direction along the junction and it is sandwiched from below and above along the $y$ direction 
by two ferromagnetic insulators ({\it F}1 and {\it F}2), forming the spin valve. 
We adopt ferromagnetic insulators as ferromagnets in the spin valve not to leak the current in the {\it N} to the ferromagnets. 
We assume that the interfaces between {\it S} and {\it N} are resistive with a finite resistance and thus low transparent. 
The similar Josephson junctions with a spin valve composed of two {\it FM}s have been previously 
studied~\cite{karminskaya-2,karminskaya-3,karminskaya-4}.

In the diffusive transport limit with the resistive interfaces, the Josephson current is evaluated by solving the linearized Usadel equation 
in the {\it N}~\cite{golubov-rmp, buzdin-rmp, bergeret-rmp, demler-prb}, 
\begin{eqnarray}
\hbar D \partial _{x}^{2} f_{\pm }(x) 
&-&2 \hbar |\omega| f_{\pm }(x)  
\pm i 2 H_{\rm eff} f_{\pm }(x) \nonumber \\
&\mp& \frac{2\hbar}{\tau_{\rm SO}}
\left[
f_{\pm }(x) - f_{\mp }(x)
\right]
=0,
\label{usadel}
\end{eqnarray}
where {\it D} and $\tau_{\rm SO}$ are the diffusion coefficient and the relaxation time due to the spin-orbit scattering, respectively, in the {\it N} 
and $\omega$ is the fermion Matsubara frequency. 
The anomalous Green's function $f_{\uparrow\downarrow}(x)$ [$f_{\downarrow\uparrow}(x)$] for spin up (down) and down (up) electrons 
in the {\it N} is denoted simply by $f_+(x)$ [$f_-(x)$]. 
We also assume that the thickness of {\it N} in the $y$ direction perpendicular to the junction is very thin (a few--10 nm), in which 
the effective exchange field $H_{\rm eff}$ inside the {\it N} induced by the MPE can be approximated to be uniform, i.e., 
\begin{eqnarray}
{H_{\rm eff}} = \left\{ \begin{array}{l}
{h_{\rm eff1}} + {h_{\rm eff2}},\,\,\,\,\ {\rm for\, P} \\
{h_{\rm eff1}} - {h_{\rm eff2}}{\rm{,\,\,\,\,\ {\rm for\, AP}}} 
\end{array} \right.\
\end{eqnarray}
for parallel (P) and antiparallel (AP) magnetization configuration between {\it F}1 and {\it F}2 (see Fig.~\ref{sns-sv}), where 
$h_{\rm eff1}$ ($h_{\rm eff2}$) is an effective exchange field induced by the MPE due to the 
magnetization in {\it F}1 ({\it F}2). 
Here, we consider the case where the magnetization direction of {\it F}1 is fixed by exchange bias, while the magnetization direction 
of {\it F}2 is freely flipped by an external magnetic field~\cite{handbook,note1}. 

As appropriate boundary conditions for the resistive interfaces between {\it S} and {\it N}, 
we impose Kupriyanov-Lukichev boundary condition~\cite{kuprianov}, 
\begin{eqnarray}
f_{\pm }^{\rm S}|_{x=-d_{\rm N}} &=& f_{\pm}(x)|_{x=-d_{\rm N}} - \xi_{\rm D} \gamma_{\rm B} \partial _{x}  f_{\pm}(x)|_{x=-d_{\rm N}}
\label{klbc1}, \\
f_{\pm }^{\rm S}|_{x=d_{\rm N}} &=& f_{\pm}(x)|_{x=d_{\rm N}} + \xi_{\rm D} \gamma_{\rm B} \partial _{x}  f_{\pm}(x)|_{x=d_{\rm N}}
\label{klbc2}, 
\end{eqnarray}
where $f_{\pm }^{\rm S}|_{x=(-)d_{\rm N}}=\mp i\Delta  e^{i\theta_{\rm R (L)}} /\sqrt{(\hbar \omega)^{2} + \Delta^{2}}$ is 
the anomalous Green's function in the S, attached to the right (left)  side of the {\it N}, with the superconducting gap $\Delta$ and the superconducting 
phase $\theta_{\rm R(L)}$. $\xi_{\rm D}=\sqrt{\hbar D/2 \pi k_{\rm B} T_{\rm C}}$ and  
$T_{\rm C}$ is the superconducting transition temperature. 
The dimensionless parameter $\gamma_{\rm B}=R_{\rm b}\sigma_{\rm N}/\xi_{\rm D}$ 
is the {\it S}/{\it N} interface transparency parameter~\cite{buzdin-rmp, bergeret-rmp}, 
where $R_{\rm b}$ is the interface resistance per unit area 
and $\sigma_{\rm N}$ is the conductivity of {\it N}. 
The interface is resistive when $\gamma_{\rm B}\gg 1$. 
Solving Eq.~(\ref{usadel}) with these boundary conditions, the anomalous Green's function in the {\it N} is given as

\begin{eqnarray}
\left( \begin{array}{l}
f_{ + }(x)\\
f_{ - }(x)
\end{array} \right) &=& \frac{{\left( {\alpha  + i} \right)\Delta }}{{1 - {\alpha ^2}}}\frac{1}{ {  {\Phi _\omega }\left( {{d_{\rm N}}} \right)}  \sqrt { {{\left( {\hbar {\omega}} \right)}^2} 
+ {\Delta ^2}}  }   \left( \begin{array}{l}
i\alpha \\
1
\end{array} \right)\left[ {{e^{i{\theta _{\rm{R}}}}}{\chi _ + }\left( {x + {d_{\rm N}}} \right) - {e^{i{\theta _{\rm{L}}}}}{\chi _ - }\left( {x - {d_{\rm N}}} \right)} \right]  \nonumber \\
&+& \frac{{\left( {\alpha  - i} \right)\Delta }}{{1 - {\alpha ^2}}}\frac{1}{{ \Phi _\omega ^*\left( {{d_{\rm N}} } \right) \sqrt {{{\left( {\hbar {\omega}} \right)}^2} 
+ {\Delta ^2}}  } }\left( \begin{array}{l}
1\\
 - i\alpha 
\end{array} \right)\left[ {{e^{i{\theta _{\rm{R}}}}}\chi _ + ^*\left( {x + {d_{\rm N}}} \right) - {e^{i{\theta _{\rm{L}}}}}\chi _ - ^*\left( {x - {d_{\rm N}}} \right)} \right], 
%
\label{fpm}
\end{eqnarray}
where
\begin{eqnarray}
{\chi _ \pm }\left( {x \pm {d_{\rm N}}} \right) &=& \sinh \left[ {{\kappa _{\rm N}}\left( {x \pm {d_{\rm N}}} \right)} \right] \nonumber \\
&\pm& {\xi _{\rm D}}{\gamma _{\rm{B}}}{\kappa _{\rm N}}\cosh \left[ {{\kappa _{\rm N}}\left( {x \pm {d_{\rm N}}} \right)} \right] 
\end{eqnarray}
and
\begin{eqnarray}
\Phi_{\omega}(d_{\rm N}) &=& 2{\xi _{\rm D}}{\gamma _{\rm{B}}}{\kappa _{\rm N}}\cosh \left( {2{\kappa _{\rm N}}{d_{\rm N}}} \right) \nonumber\\
&+& \left[ {1 + {{\left( {{\xi _{\rm D}}{\gamma _{\rm{B}}}{\kappa _{\rm N}}} \right)}^2}} \right]\sinh \left( {2{\kappa _{\rm N}}{d_{\rm N}}} \right) 
\end{eqnarray}
with 
\begin{equation}
\kappa_{\rm N}=\sqrt{ \frac{2 \hbar|\omega| + 2\hbar/\tau_{\rm SO} }{ \hbar D } - i2 \sqrt{ \frac{H_{\rm eff}^{2} - (\hbar/\tau_{\rm SO})^2 }{(\hbar D)^2} } } 
\end{equation}
and 
\begin{equation}
\alpha=\frac{ \hbar/\tau_{\rm SO} }{ \sqrt{H_{\rm eff}^{2} - (\hbar/\tau_{\rm SO})^2} + H_{\rm eff} }. 
\end{equation}

Within the quasiclassical theory and the linearized approximation, the Josephson current density flowing through the {\it N} is 
given as \cite{buzdin-rmp, bergeret-rmp}
\begin{eqnarray}
j_{\rm{J}}\left( {{d_{\rm N}}}, \theta \right) &=& 
\frac{{2 \pi e{N_{\rm F}} D}}{\beta }
\sum\limits_{\omega>0 } {\mathop{\rm Im}\nolimits} 
\left[ {f_ - }(x){\partial _x}f_ + ^*(x)  \right. \nonumber \\
&{}&  \quad\quad \quad\quad\quad\quad\quad     \left. + {f_ + }(x){\partial _x}f_ - ^*(x)  \right],
\label{jj}
\end{eqnarray}
where $N_{\rm F}$ is the density of states per unit volume and per electron spin at the Fermi energy~\cite{note4}, 
$\beta=1/k_{\rm B}T$ ($T$: temperature), and $\theta=\theta_{\rm R}-\theta_{\rm L}$ is the phase difference between 
the right and left Ss. Substituting Eq.~(\ref{fpm}) into Eq.~(\ref{jj}), we obtain the Josephson current density 
\begin{equation}
j_{\rm J}(d_{\rm N}, \theta)=j_{\rm c}(d_{\rm N}) \sin\theta, 
\end{equation}
where  
\begin{equation}
j_{\rm c}(d_{\rm N}) = 
	-\frac{4 \pi e N_{\rm F} D \Delta^{2} }{\beta}
	\sum_{\omega>0} {\rm Re}
	\left[
	{\rm Re}(A) \kappa_{\rm N} \Psi_{\omega}(d_{\rm N})
	\right]
\label{jc1}
\end{equation}
%
with
\begin{eqnarray}
&{}&\Psi_{\omega}(d_{\rm N}) = (\alpha + i) (\alpha^{*} + i)\frac{2\xi_{\rm D} \gamma_{\rm B} \kappa_{\rm N} \cosh (2 \kappa_{\rm N} d_{\rm N} ) }
	{[(\hbar \omega )^{2}+\Delta^{2}] \Phi_{\omega}^{2}(d_{\rm N})} \nonumber \\
 &{}& +  (\alpha + i) (\alpha^{*} + i)\frac{[1+(\xi_{\rm D} \gamma_{\rm B} \kappa_{\rm N})^2] \sinh(2 \kappa_{\rm N} d_{\rm N} )}
	{[(\hbar \omega )^{2}+\Delta^{2}] \Phi_{\omega}^{2} (d_{\rm N})} 
\end{eqnarray}
and 
\begin{equation}
A=\frac{1}{1-{\alpha^*}^2}. 
\end{equation}
Notice that $j_{\rm c} (d_{\rm N})$ is the Josephson critical current density 
and we call it simply Josephson critical current below. 


\section{Results}\label{sec:result}

\begin{figure}[t!]
\begin{center}
\includegraphics[width=7cm]{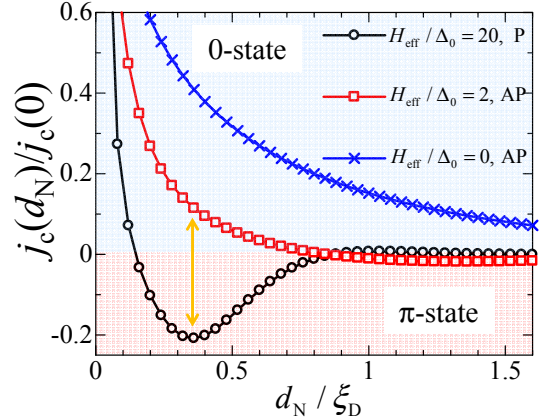}
\caption{ (color online) The Josephson critical current $j_{\rm c}$($d_{\rm N}$) as a function of $d_{\rm N}$. 
Here, we set $T/T_{\rm C}=0.3$, $\gamma_{\rm B}=10$~\cite{note5}, and $\hbar/\tau_{\rm SO} =0$. 
$\Delta_{0}$ is the superconducting gap at zero temperature and $\xi_{\rm D}=\sqrt{\hbar D/2 \pi k_{\rm B} T_{\rm C}}$. 
P (AP) denotes parallel (antiparallel) magnetization configuration between {\it F}1 and {\it F}2 in the spin valve. 
As indicated by the vertical arrow, the Josephson junction changes from $\pi$-state to 0-state 
(and vice versa) by varying the magnetization configuration from P to AP (from AP to P) when 
$d_{\rm N}$ is set to be about $0.35\,\xi_{\rm D}$. 
}
\label{jc-d}
\end{center}
\end{figure}

Let us first numerically evaluate $j_{\rm c} (d_{\rm N})$ in the {\it N} by using Eqs.~(\ref{jc1}). 
For this purpose, the temperature dependence of $\Delta$ is assumed to be $\Delta=\Delta_{0} \tanh(1.74\sqrt{T_{\rm C}/T-1})$, 
where $\Delta_{0}$ is the superconducting gap at zero temperature~\cite{yamashita-prb}. 
Figure~\ref{jc-d} shows the representative result of the Josephson critical current $j_{\rm c}(d_{\rm N})$ as a function of $d_{\rm N}$ 
for different $H_{\rm eff}$ without spin-orbit scattering. 
First, it is found in Fig.~\ref{jc-d} (indicated by circles) that 
$j_{\rm c}(d_{\rm N})$ clearly exhibits damped oscillatory behavior with increasing $d_{\rm N}$ 
when the magnetization configuration between {\it F}1 and {\it F}2 is parallel, revealing the 0-$\pi$ transition with $d_{\rm N}$. 
Secondly, it is also found that the partial cancellation of effective exchange fields $h_{{\rm eff}1}$ and $h_{{\rm eff}2}$, 
as a result of the AP magnetization 
configuration between {\it F}1 and {\it F}2, remarkably transmutes the period of oscillation and the damping rate of 
$j_{\rm c}(d_{\rm N})$ (denoted by squares in Fig.~\ref{jc-d}). Thirdly, $j_{\rm c}(d_{\rm N})$ eventually decreases monotonically 
with increasing $d_{\rm N}$, as shown by crosses in Fig.~\ref{jc-d}, when the effective exchange fields are perfectly canceled 
for the AP magnetization configuration. 
This monotonic decrease of $j_{\rm c}(d_{\rm N})$ with $d_{\rm N}$ is the same as that in {\it S}/{\it N}/{\it S} junctions. 

These results clearly demonstrate that the {\it S}/{\it N}/{\it S} junction with the spin valve can control reversibly 
the 0-$\pi$ transition simply by changing the magnetization direction between {\it F}1 and {\it F}2 when the thickness of {\it N} is set 
appropriately, for instance, about $0.35\, \xi_{\rm D}$ for the case shown in Fig~\ref{jc-d} (indicated by the vertical arrow). 
This 0-$\pi$ transition is driven by the effective exchange field in the {\it N} induced by the MPE between {\it F} and {\it N} and therefore 
it is sharply contrasted to the previously reported one based on {\it S}/{\it N}/{\it S} junctions operated by the DC 
voltage~\cite{volkov-prl74, wilhelm-prl81, yip-prb58, baselmans-nature, baselmans-aps}.

\begin{figure}[!t]
\begin{center}
\includegraphics[width=7cm]{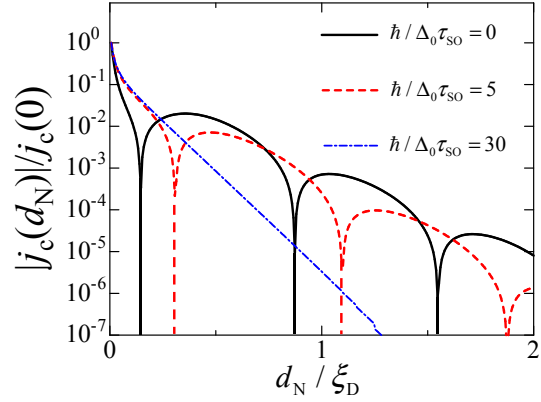}
\caption{ (color online) The absolute value of Josephson critical current $j_{\rm c}$($d_{\rm N}$) as a function of $d_{\rm N}$ 
for $\hbar/\Delta_{0}\tau_{\rm SO}=0$, $5$, and $30$. Here, $\Delta_{0}$ is the superconducting gap at zero temperature 
and $\xi_{\rm D}=\sqrt{\hbar D/2 \pi k_{\rm B} T_{\rm C}}$. 
The other parameters are $H_{\rm eff}/\Delta_0=20$, $T/T_{\rm C}=0.3$, and $\gamma_{\rm B}=10$~\cite{note5}. 
The period of oscillation and the damping rate of $j_{\rm c}$($d_{\rm N}$) are affected sensitively by $\tau_{\rm SO}$. 
}
\label{jc-d-soi}
\end{center}
\end{figure}

\begin{figure}[t!]
\begin{center}
\vspace{30mm} 
\includegraphics[width=6.5cm]{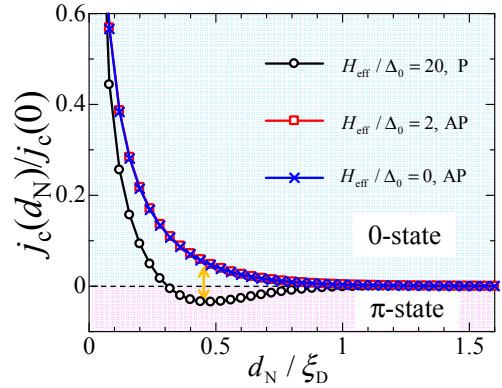}
\caption{ (color online) Same as Fig.~\ref{jc-d} but with $\hbar/\Delta_{0}\tau_{\rm SO} =5$. 
As indicated by the vertical arrow, the Josephson junction changes from $\pi$-state to 0-state 
(and vice versa) by varying the magnetization configuration from P to AP (from AP to P) when 
$d_{\rm N}$ is set to be about $0.45\,\xi_{\rm D}$. 
}
\label{jc-d2}
\end{center}
\end{figure}

Let us now evaluate numerically the $\tau_{\rm SO}$ dependence of the Josephson critical current 
$j_{\rm c}(d_{\rm N})$~\cite{demler-prb}. 
Figure~\ref{jc-d-soi} shows $j_{\rm c}(d_{\rm N})$ as a function of $d_{\rm N}$ for three different values of $\tau_{\rm SO}$. 
As seen in Fig.~\ref{jc-d-soi}, 
the period of oscillation with $d_{\rm N}$ is longer and the damping rate is stronger as $\tau_{\rm SO}$ decreases. 
Further decrease of $\tau_{\rm SO}$ diminishes completely the oscillatory behavior and eventually 
$j_{\rm c}(d_{\rm N})$ decreases monotonically as a function of $d_{\rm N}$, as indicated by dot-dashed line in 
Fig.~\ref{jc-d-soi}. 
Therefore, the period of oscillation and the damping rate of $j_{\rm c}(d_{\rm N})$ are rather sensitive to 
the spin-orbit scattering. However, for a reasonable value of $\tau_{\rm SO} $ 
(see below in Sec.~\ref{sec:discussion}), the 0-$\pi$ transition can be still reversibly controlled by varying the magnetization 
direction between {\it F}1 and {\it F}2 when the thickness of {\it N} is set appropriately, as shown in Fig~\ref{jc-d2}.

Next, we shall discuss the $\tau_{\rm SO}$ dependence of $j_{\rm c}(d_{\rm N})$ by considering approximated formula 
of Eq.~(\ref{jc1}). 
For $T\approx T_{\rm C}$ and $d_{\rm N}/\xi_{\rm D} \gg 1$, $j_{\rm c}(d_{\rm N})$ is approximated as 
\begin{eqnarray}
j_{\rm c}(d_{\rm N}) &\propto& 
	\left \{
	{\rm Re}(j_0)
	\cos [2{\rm Im}( \kappa_{N}) d_{\rm N} ] \right. \nonumber \\
	&+&  \left. 
	{\rm Im}(j_0)
	\sin [2{\rm Im}( \kappa_{N} ) d_{\rm N} ] 
	\right \} e^{-2{\rm Re}(\kappa_{\rm N}) d_{\rm N}} 
\label{jc-ap} 
\end{eqnarray}
with 
$j_0= (\alpha+i ) ( \alpha^{*} + i) / \left(\xi_{\rm D}^2 \gamma_{\rm B}^2 \kappa_{\rm N} \right)$. 
Therefore, the oscillation of $j_{\rm c}(d_{\rm N})$ as a function of $d_{\rm N}$ is determined solely by the imaginary part of 
$\kappa_{\rm N}$. 
We now consider the two limiting cases for $\kappa_{\rm N}$. 
The first limiting case is when $H_{\rm eff}$ is much larger than $\hbar/\tau_{\rm SO}$. 
In this limit, $\kappa_{\rm N}\approx \sqrt{ \frac{2\pi k_{\rm B} T - i 2 |H_{\rm eff}|}{\hbar D} }$.
Therefore, $j_{\rm c}(d_{\rm N})$ shows the damped oscillatory behavior with $d_{\rm N}$ because 
${\rm Im}(\kappa_{\rm N}) \neq 0 $. 
The other limiting case is when $H_{\rm eff}$ is much smaller than $\hbar/\tau_{\rm SO}$. 
In this limit, $\kappa_{\rm N}\approx \sqrt{ \frac{ 2\pi k_{\rm B} T + 4\hbar/\tau_{\rm SO} }{ \hbar D } }$
and thus ${\rm Im}(\kappa_{\rm N}) = 0 $. 
Therefore, $j_{\rm c}(d_{\rm N})$ decreases monotonically with $d_{\rm N}$ because the oscillating terms in Eq.~(\ref{jc-ap}) 
are absent. These results are indeed comparable with the numerical results shown in Figs.~\ref{jc-d-soi} and \ref{jc-d2}.

\section{Discussion}\label{sec:discussion}

We shall now approximately estimate $\tau_{\rm SO}$, $H_{\rm eff}$, and $\xi_{\rm D}$ for realistic systems. 
As an example, we consider Pt for the {\it N} and ${\rm Y_3Fe_2(FeO_4)_3}$ (YIG) for the {\it F}s, for which $\hbar/\tau_{\rm SO}$ and 
$H_{\rm eff}$ for P are roughly 
estimated to be about 5 meV and 20 meV, respectively~\cite{lu-prl110, wilhelm-prl85, robinson, takahashi, kimura-prl98}. 
Thus, $\hbar/\Delta_0\tau_{\rm SO}\approx 5$ and $H_{\rm eff}/\Delta_0\approx20$ for P when we take 
$\Delta_{0}=1$ meV for Nb as the {\it S}s~\cite{nb-para}, 
which corresponds to the red dashed line and open circles in Figs.~\ref{jc-d-soi} and \ref{jc-d2}, respectively. 
In the case of Pt, $\xi_{\rm D}$ is estimated to be about 20 nm~\cite{lu-prb87, lu-prl110, note6, book-ssp}, and hence the thickness of 
{\it N} for which the junction exhibits the 0-$\pi$ transition by varying the magnetization direction between {\it F}1 and {\it F}2 in the spin valve 
is the order of several dozen nanometers (see Fig.~\ref{jc-d2}).
In this case, the Josephson critical current is approximately estimated to be one to two orders of 
magnitude smaller than  
\begin{equation}
j_c(0)\approx \frac{4\pi e N_{\rm F} D k_{\rm B}T_{\rm C}}{\xi_{\rm D}}\times 3\times10^{-3}.
\end{equation}
Assuming the free density of states for Pt~\cite{note7}, $j_c(0)\approx1\times10^9\, {\rm A/m^2}$~\cite{note6}. 
Thus, the Josephson critical current can be as large as the one reported in Ref.~\cite{robinson}. 
Note also that the direction of the ferromagnetic magnetization in YIG is easily flipped by applying an external magnetic field as 
large as hundreds of oersteds~\cite{handbook}.  
Therefore, we expect that the 0-$\pi$ transition can be 
experimentally observed within the current nanofabrication techniques if Pt is used for the {\it N} attached to YIG in the spin valve~\cite{note8}. 

\section{Summary}\label{sec:summary}

We have studied the Josephson effect in the {\it S}/{\it N}/{\it S} junction composed of $s$-wave {\it S}s with the {\it N} sandwiched 
by two {\it F}s in the vertical direction of the junction, forming the spin valve. 
We have shown that the 0-$\pi$ transition is driven by the effective exchange field in the {\it N} which is 
induced by the MPE in the spin valve. 
Moreover, we have shown that the 0- and $\pi$-states of this Josephson junction are reversibly switchable when 
the magnetization configuration of two {\it F}s are changed from parallel to antiparallel configuration and vice versa. 
We have also found that the spin-orbit scattering influences the Josephson critical current rather sensitively. Finally, we have 
estimated the Josephson critical current for realistic systems and found that the 0-$\pi$ transition can be observed in the currently 
available experimental setup, indicating a promising potential of this junction as a 0-$\pi$ switching device operated by an external 
magnetic field. 
Our results not only provide possible applications in superconducting electronics but also suggest the importance of a 
fundamental concept of MPE in nanostructures of multilayer ferromagnetic systems

\section*{ACKNOWLEDGMENTS} 

This work is supported by Grant-in-Aid for Research Activity Start-up (No. 25887053) from the Japan Society for the
Promotion of Science and also in part by RIKEN iTHES Project.               

{}

\begin{thebibliography}{}
\bibitem{note1}
The external magnetic field required for the magnetization reversal in the spin valve is 
within a few hundred of oersteds~\cite{handbook} and thus it is not large enough to break superconductivity in e.g., Nb~\cite{nb-para}.
\bibitem{note4}
In the quasiclassical Green's function theory, 
$N_{\rm F}$ for up and down electrons in the spin polarized N due to the MPE is assumed to be 
approximately the same~\cite{golubov-rmp,buzdin-rmp,bergeret-rmp}.
\bibitem{note5}
The value of $\gamma_{\rm B}$ varies with 
the resistance of the junction, conductivity of N, and $\xi_{\rm D}$ 
in the N~\cite{buzdin-rmp}. 
For example, $\gamma_{\rm B}$ is estimated to be 5--10 for a Nb/PdNi junction~\cite{kontos-prl}.
\bibitem{note7}
$N_{\rm F}=\frac{1}{4\pi^2}\left( \frac{2m}{\hbar^2} \right)^{3/2}\varepsilon_{\rm F}^{1/2}$ per spin with the Fermi energy 
$\varepsilon_{\rm F}\approx1$ eV, where $m$ is the electron mass.
\bibitem{note8}
On the other hand, when Cu or Au is used for the N attached to YIG, the induced effective exchange field is 
negligibly small~\cite{huang-prl109,qu-prl110}. Thus, our calculations predict that the 0-$\pi$ transition can not be observed in this case 
(see Figs.~\ref{jc-d} and \ref{jc-d2}).
\end{thebibliography}
\end{document}